\documentclass[%
preprint,
amsmath,amssymb,
prb,
]{revtex4-2}

\usepackage{graphicx}% Include figure files
\usepackage{dcolumn}% Align table columns on decimal point
\usepackage{bm}% bold math
\usepackage{hyperref}% add hypertext capabilities
\usepackage{float}
\usepackage{orcidlink}
\newcommand{\iitj}{\affiliation{Department of Physics, Indian Institute of Technology Jodhpur\\
		N.H. 62, Nagaur Road, Karwar, Jodhpur, Rajasthan, India - 342030.}}

\begin{document}
	\newcommand{\revsapta}[1]{{\color{red}{#1}}}
	\preprint{} 
	
	\title{Extreme Suppression of Thermal Conductivity in Monolayer Beryllene}% Force line breaks with \\
	
	\author{Sapta Sindhu Paul Chowdhury$\,$\orcidlink{0009-0007-0472-7660}}
	\email{saptasindhu62@gmail.com}
	%\sag
	\iitj
	
	%\altaffiliation[Also at ]{Scientific Analysis Group, Metcalfe House, Civil Lines, Delhi, India - 110054}%Lines break automatically or can be forced with \\
	\author{Santosh Mogurampelly$\,$\orcidlink{0000-0002-3145-4377}}%
	\email{santosh@iitj.ac.in}
	\iitj
	%\affiliation{%
		% Department of Physics, Indian Institute of Technology Jodhpur\\
		% N.H. 62, Nagaur Road, Karwar, Jodhpur, Rajasthan, India - 342030. 
		%}%
	
	
	%\affiliation{
	%\affiliation{
	%\affiliation{%
	
	
	\date{\today}% It is always \today, today,
	
	\begin{abstract}
    We demonstrate an extreme suppression of thermal conductivity in monolayer beryllene through first-principles solutions of the phonon Boltzmann transport equation. While the fully converged three-phonon calculation predicts a room-temperature lattice thermal conductivity of 154.3 W/(m$\cdot$K), inclusion of four-phonon interactions suppresses it to 10.7 W/(m$\cdot$K), a reduction of more than 90\%. This exceptionally large reduction of the thermal conductivity is amongst the highest reported in two-dimensional materials. The microscopic origin of this breakdown is traced to the strong sensitivity of the low-frequency flexural ZA phonons to four-phonon scattering: mode-resolved scattering rates reveal a pronounced enhancement of ZA scattering, which substantially shortens their lifetimes and suppresses their dominant contribution to heat conduction. Moreover, the temperature dependence of the thermal conductivity changes from $\kappa\propto T^{-0.86}$ to $\kappa\propto T^{-1.38}$ upon inclusion of four-phonon processes. Our results establish $\alpha$-beryllene as an unusually strong case of four-phonon-controlled thermal transport amongst two-dimensional materials. These findings identify selective higher-order scattering of flexural phonons as a key mechanism for controlling thermal transport in strongly anharmonic low-dimensional systems.
	\end{abstract}
	
	\maketitle
	
	%\tableofcontents
	
	\section{\label{sec:introduction}Introduction\protect}%\\ %The line
	%break was forced \lowercase{via} \textbackslash\textbackslash
	The discovery of graphene has spurred intense research into two-dimensional (2D) materials due to their extraordinary electronic, optical, mechanical, and thermal characteristics \cite{Novoselov2004, Novoselov2019}. This breakthrough has subsequently led to the emergence of a broader class of atomically thin systems, including silicene, germanene, stanene, phosphorene, borophene, and other elemental Xenes, each exhibiting a rich spectrum of physical phenomena rooted in reduced dimensionality, quantum confinement, and diverse bonding configurations \cite{Molle2025,Grazianetti2024}. 
	
	Among different physical properties of these materials, thermal transport has attracted particular attention because efficient heat dissipation is essential for the reliable operation of nanoelectronic, optoelectronic, thermoelectric, and quantum devices \cite{Grazianetti2026}. Consequently, understanding the microscopic mechanisms of phonon transport has become a central theme in the study of low-dimensional materials. Graphene, with its exceptionally high room-temperature thermal conductivity, has served as the benchmark for phonon-mediated heat transport \cite{Lindsay2014, Fugallo2014,Balandin2008}. Subsequent studies on silicene, germanene, stanene, phosphorene, and borophene have demonstrated how lattice buckling, bonding characteristics, and anharmonicity profoundly influence thermal conductivity \cite{Gu2018,Sang2019, sapta2023, sapta_germanene}.
	
	Among the recently discovered elemental monolayers, $\alpha$-beryllene has emerged as a particularly intriguing member of the Xene family. This lightest 2D material was first predicted by Sun \textit{et al.} using first-principles calculations. The $\alpha$- and $\beta$-beryllene were shown to be dynamically and thermodynamically stable metallic monolayers possessing unusual planar multicenter bonding and excellent thermal stability \cite{beryllene_sun20}. More recently, Chahal \textit{et al.} experimentally synthesized free-standing beryllene nanosheets through a sonochemical route, confirming the existence of atomically thin beryllium with metallic conductivity and revealing the coexistence of multiple crystalline polymorphs \cite{Chahal2023}. As the lightest experimentally realized elemental Xene, beryllene combines an ultralow atomic mass with a perfectly planar lattice, making it fundamentally distinct from graphene and the buckled group-IV monolayers.
	
	The successful realization of beryllene has stimulated rapidly growing interest in its fundamental properties and technological applications. Owing to its metallic electronic structure, high theoretical specific capacity, low diffusion barriers, and structural stability, $\alpha$-beryllene has been proposed as a promising anode material for sodium- and potassium-ion batteries \cite{beryllene_sun20}. Its electrochemical potential has subsequently been extended to multivalent energy-storage systems, where first-principles calculations predict excellent magnesium-ion storage performance with fast ion diffusion and high reversible capacity \cite{Gao2023}. Beyond electrochemical applications, recent investigations have revealed an equally rich electronic landscape. Li \textit{et al.} predicted that both $\alpha$- and $\beta$-beryllene exhibit phonon-mediated superconductivity arising from strong electron-phonon coupling associated with their in-plane covalent bonding and characteristic vibrational modes \cite{Li2023}. More recently, Jiang \textit{et al.} demonstrated that beryllene simultaneously hosts non-trivial topological electronic states and superconductivity, an unusual coexistence among elemental two-dimensional materials that further shows its potential as a quantum material \cite{Jiang2025}. In addition, Ghavami \textit{et al.} theoretically investigated a beryllene/graphene/beryllene trilayer heterostructure and showed electrically tunable plasmon dispersion together with transitions from non-convex to hyperbolic plasmonic regimes, suggesting promising applications in plasmonics and nanoscale optoelectronic devices \cite{GHAVAMI2024}. Recent theoretical studies have also explored the structural polymorphism of atomically thin beryllium, further enriching the growing family of stable two-dimensional beryllium allotropes \cite{Zhang2025be}.
	
	Despite these significant advances in understanding the structural, electronic, electrochemical, and superconducting properties of beryllene, its intrinsic lattice thermal transport remains comparatively unexplored. Understanding thermal transport in $\alpha$-beryllene is critical because nearly every proposed application of beryllene, including nanoelectronics, superconducting devices, battery electrodes, and plasmonic heterostructures, requires efficient heat dissipation for stable operation. Moreover, thermal conductivity provides direct insight into the anharmonic lattice dynamics, phonon scattering mechanisms, and symmetry-governed heat transport that cannot be inferred solely from electronic structure calculations. Unlike graphene, whose exceptional thermal conductivity originates from strong carbon-carbon covalent bonding, or buckled Xenes such as silicene and germanene, where enhanced anharmonicity suppresses phonon transport, $\alpha$-beryllene combines metallic bonding, an ultralow atomic mass, and a perfectly planar lattice. Bulk beryllium itself possesses one of the highest thermal conductivities among elemental metals, together with a high Debye temperature, low specific mass, and strong interatomic bonding, naturally raising the question of how these favorable thermal transport characteristics evolve in its atomically thin counterpart \cite{ginningsJACS1951,hidnert-sweeneyNIST, Powell1953}.

	To address this fundamental gap, we present a comprehensive first-principles investigation of the intrinsic lattice thermal conductivity of $\alpha$-beryllene by an iterative solution of the phonon Boltzmann transport equation. Because strictly planar, ultralight 2D lattices often exhibit highly anomalous flexural mode behavior, we explicitly include both three- and four-phonon scattering processes to capture the complete transport picture \cite{Lindsay2010, bonini2012acoustic, Feng2016, Feng2017}. Beyond determining the magnitude and temperature dependence of the thermal conductivity, we explain the microscopic mechanisms governing heat transport through detailed analyses of phonon dispersions, mode-resolved thermal conductivity, phonon lifetimes, scattering phase space, and spectral accumulation. We show that including four phonon scatterings in $\alpha$-beryllene reduces the room-temperature lattice thermal conductivity from 154.3 W/(m$\cdot$K) to 10.7 W/(m$\cdot$K), a reduction of about 93\%. 

	\section{\label{sec:theory}Theoretical Formalism and Computational Details\protect}%\\ %The line
	%break was forced \lowercase{via} \textbackslash\textbackslash}

Density functional theory (DFT) calculations were performed using the \textsc{Quantum ESPRESSO} package \cite{Giannozzi2007,Giannozzi2017}. The exchange-correlation interaction was described within the generalized gradient approximation (GGA) using the Perdew-Burke-Ernzerhof (PBE) functional \cite{pbe1996}. Electron-ion interactions were treated using norm-conserving pseudopotentials from the precision dataset of the Standard Solid-State Pseudopotential (SSSP) library \cite{sssp2018}. An optimized plane-wave kinetic energy cutoff of 80 Ry (Fig. S1) with a charge-density cutoff corresponding to a dual ratio of 4 ($E_{\rm cut}^{\rho}/E_{\rm cut}^{\rm wfc}=4$) was employed throughout the calculations.

The atomic structure was fully relaxed until the Hellmann-Feynman forces on each atom were smaller than $10^{-8}$ eV/\AA~and the total energy was converged to within $10^{-10}$ eV. Owing to the metallic nature of $\alpha$-beryllene, Brillouin-zone integrations were performed using a dense Monkhorst-Pack \cite{mp1976} $72\times72\times1$ \textit{k}-point mesh together with the Marzari-Vanderbilt cold smearing scheme \cite{mvcold} employing a smearing width of 0.01 eV. A vacuum layer of 25~\AA~was introduced along the out-of-plane direction to eliminate spurious interactions between periodically repeated images.

The harmonic interatomic force constants (IFCs) were calculated using density functional perturbation theory (DFPT), which is based on the $2n+1$ theorem \cite{Baroni2001}. Although harmonic IFCs may also be obtained using the finite-displacement method, the acoustic phonon branches generally exhibit slower convergence with respect to supercell size in finite-displacement calculations, particularly for two-dimensional materials \cite{dfptvsfd2019}. After comparing both approaches, DFPT was adopted owing to its superior accuracy for describing the low-frequency acoustic phonons. The dynamical matrices were computed on a converged $36\times36\times1$ \textit{q}-point mesh and subsequently used for all lattice thermal conductivity calculations.

The third-order anharmonic IFCs were obtained using the finite-displacement approach implemented in \textsc{Thirdorder.py} \cite{thirdorder2012}. A $16\times16\times1$ supercell was employed while including interactions up to the eleventh nearest-neighbor shell. To account for higher-order anharmonicity, the fourth-order IFCs were computed using the \textsc{Fourthorder.py} package \cite{Han2022,Feng2016} with the same $16\times16\times1$ supercell by considering interactions up to the second nearest-neighbor shell. These interaction cutoffs were adopted following convergence tests of the calculated lattice thermal conductivity (Fig. S2). Computing four phonon scattering processes is resource-intensive. So, a sampling-accelerated method for the phonon scattering rate is used to address this issue \cite{Guo2024, Guo2023}. An optimized sampling size of 3500000 is used for sampling four phonon scattering events (Fig. S3).

The lattice thermal conductivity was evaluated by iteratively solving the linearized phonon Boltzmann transport equation ($l$BTE) using the \textsc{ShengBTE} package \cite{Li2014ShengBTE} on a $144\times144\times1$ \textit{q}-point mesh (Fig. S4). Within the iterative solution, the lattice thermal conductivity tensor is expressed as:
\begin{eqnarray}
	\kappa^{\alpha\beta}_{\rm L}
	&=&
	\frac{\hbar^2}
	{k_{\rm B}T^2VN_q}
	\sum_{\mathbf{q}j}
	\omega_{\mathbf{q}j}^2
	n_{\mathbf{q}j}
	\left(n_{\mathbf{q}j}+1\right)
	\nonumber\\
	&&\times
	\left(
	v_{\mathbf{q}j}^{\alpha}
	+\Delta_{\mathbf{q}j}^{\alpha}
	\right)
	\left(
	v_{\mathbf{q}j}^{\beta}
	+\Delta_{\mathbf{q}j}^{\beta}
	\right)
	\tau_{\mathbf{q}j},
	\label{eq_bte}
\end{eqnarray}
where $k_{\rm B}$ is the Boltzmann constant, $\hbar$ is the reduced Planck constant, $T$ is the absolute temperature, $V$ is the effective unit-cell volume, and $N_q$ denotes the total number of sampled \textit{q}-points in the first Brillouin zone. The phonon mode with wave vector $\mathbf{q}$ and branch index $j$ has angular frequency $\omega_{\mathbf{q}j}$, group velocity $v_{\mathbf{q}j}^{\alpha}$ along the Cartesian direction $\alpha$, lifetime $\tau_{\mathbf{q}j}$, and equilibrium Bose-Einstein occupation number:
\begin{equation}
	n_{\mathbf{q}j}
	=
	\frac{1}
	{\exp\!\left(\hbar\omega_{\mathbf{q}j}/k_{\rm B}T\right)-1}.
\end{equation}
The quantities $\Delta_{\mathbf{q}j}^{\alpha}$ and $\Delta_{\mathbf{q}j}^{\beta}$ denote the correction terms arising from the iterative solution of the $l$BTE beyond the relaxation time approximation (RTA), thereby accounting for the redistribution of nonequilibrium phonon populations due to momentum-conserving normal scattering processes. Following the conventional treatment for two-dimensional materials, the effective unit-cell volume was evaluated using an effective monolayer thickness of 3.06~\AA, corresponding to the van der Waals diameter of a Be atom.

The intrinsic phonon scattering rates arising from three-phonon interactions were evaluated using many-body perturbation theory within Fermi's golden rule. The total three-phonon scattering rate (inverse lifetime) of a phonon mode $(\mathbf{q},j)$ is given by:
\begin{eqnarray}
	\Gamma_{\mathbf{q}j}^{(3)}(T)
	&=&
	\frac{1}{\tau_{\mathbf{q}j}^{(3)}(T)}
	\nonumber\\
	&=&
	\frac{\pi}{\hbar^2N_q}
	\sum_{\mathbf{q}',j',j''}
	\left|
	\Phi^{(3)}_{\mathbf{q}j,\mathbf{q}'j',\mathbf{q}''j''}
	\right|^2
	\nonumber\\
	&&\times
	\Big[
	\left(
	1+n_{\mathbf{q}'j'}+n_{\mathbf{q}''j''}
	\right)
	\delta
	\left(
	\omega_{\mathbf{q}j}
	-
	\omega_{\mathbf{q}'j'}
	-
	\omega_{\mathbf{q}''j''}
	\right)
	\nonumber\\
	&&
	+
	2
	\left(
	n_{\mathbf{q}'j'}
	-
	n_{\mathbf{q}''j''}
	\right)
	\delta
	\left(
	\omega_{\mathbf{q}j}
	+
	\omega_{\mathbf{q}'j'}
	-
	\omega_{\mathbf{q}''j''}
	\right)
	\Big],
	\label{eq:3ph_rate}
\end{eqnarray}
where $\Phi^{(3)}$ denotes the third-order phonon scattering matrix element; the first term corresponds to phonon decay (emission) processes, while the second term describes phonon collision (absorption) processes. Momentum conservation is implicitly satisfied through:
\begin{equation}
	\mathbf{q}
	+
	\mathbf{q}'
	+
	\mathbf{q}''
	=
	\mathbf{G},
\end{equation}
where $\mathbf{G}$ is a reciprocal lattice vector.

The third-order scattering matrix elements are obtained from the third derivatives of the total energy with respect to the normal mode coordinates,
\begin{equation}
	\Phi^{(3)}_{\mathbf{q}j,\mathbf{q}'j',\mathbf{q}''j''}
	=
	\frac{1}{N_q}
	\frac{\partial^3E_{\rm tot}}
	{\partial X_{\mathbf{q}j}
		\partial X_{\mathbf{q}'j'}
		\partial X_{\mathbf{q}''j''}},
\end{equation}
where $E_{\rm tot}$ is the total energy of the crystal.

The derivatives with respect to the normal coordinates are related to the atomic displacements through:
\begin{equation}
	\frac{\partial}
	{\partial X_{\mathbf{q}j}}
	=
	\sum_{s,\alpha}
	\sqrt{
		\frac{\hbar}
		{2m_s\omega_{\mathbf{q}j}}
	}
	e^{\mathbf{q}j}_{s\alpha}
	\frac{\partial}
	{\partial u_{\mathbf{q}s\alpha}},
	\label{eq_deldelx}
\end{equation}
where $m_s$ is the mass of atom $s$, $e^{\mathbf{q}j}_{s\alpha}$ is the phonon eigenvector corresponding to the phonon mode $(\mathbf{q},j)$, and $u_{\mathbf{q}s\alpha}$ denotes the Fourier-transformed atomic displacement,
\begin{equation}
	u_{\mathbf{q}s\alpha}
	=
	\frac{1}{N_q}
	\sum_{\mathbf{R}}
	u_{\mathbf{R}s\alpha}
	e^{-i\mathbf{q}\cdot\mathbf{R}},
\end{equation}
where $u_{\mathbf{R}s\alpha}$ represents the displacement of atom $s$ along the Cartesian direction $\alpha$ in the unit cell located at lattice vector $\mathbf{R}$.

%When multiple intrinsic and extrinsic scattering mechanisms are present, the total phonon lifetime can be approximated using Matthiessen's rule,

%\begin{equation}
%	\frac{1}{\tau_{\mathbf{q}j}}
%	=
%	\frac{1}{\tau_{\mathbf{q}j}^{(3)}}
%	+
%	\frac{1}{\tau_{\mathbf{q}j}^{\rm iso}}
%	+
%	\frac{1}{\tau_{\mathbf{q}j}^{\rm b}},
%\end{equation}

%where $\tau_{\mathbf{q}j}^{(3)}$, $\tau_{\mathbf{q}j}^{\rm iso}$, and $\tau_{\mathbf{q}j}^{\rm b}$ denote the relaxation times associated with intrinsic three-phonon scattering, isotope scattering, and boundary scattering, respectively. In the present work, only intrinsic phonon--phonon scattering is considered unless otherwise stated.

The weighted scattering phase space provides a quantitative measure of the number of energy- and momentum-conserving scattering channels available to a phonon mode while accounting for the corresponding Bose-Einstein occupation factors. Although the weighted phase space does not explicitly include the anharmonic interaction strengths, it serves as a useful indicator of the intrinsic probability for phonon-phonon scattering and is therefore widely employed to analyze thermal transport in crystalline solids \cite{Li2014ShengBTE,Togo2015}.

For three-phonon processes, the weighted phase space is expressed as:
\begin{eqnarray}
	W^{\pm}_{\mathbf{q}j}
	&=&
	\frac{1}{2N_q}
	\sum_{\mathbf{q}',j',j''}
	\begin{Bmatrix}
		2\left(
		n_{\mathbf{q}'j'}
		-
		n_{\mathbf{q}''j''}
		\right)
		\\
		1+n_{\mathbf{q}'j'}+n_{\mathbf{q}''j''}
	\end{Bmatrix}
	\nonumber\\
	&&\times
	\frac{
		\delta
		\left(
		\omega_{\mathbf{q}j}
		\pm
		\omega_{\mathbf{q}'j'}
		-
		\omega_{\mathbf{q}''j''}
		\right)
	}
	{
		\omega_{\mathbf{q}j}
		\omega_{\mathbf{q}'j'}
		\omega_{\mathbf{q}''j''}
	},
	\label{eq:wps}
\end{eqnarray}
where the superscripts $(+)$ and $(-)$ denote the absorption and emission scattering processes, respectively. The total weighted phase space is obtained from the sum of the two contributions,
\begin{equation}
	W_{\mathbf{q}j}
	=
	W_{\mathbf{q}j}^{+}
	+
	W_{\mathbf{q}j}^{-}.
\end{equation}
The weighted phase space characterizes the available scattering channels that satisfy both energy and crystal momentum conservation and therefore provides valuable insight into the frequency dependence of intrinsic phonon scattering. However, the phonon scattering rate is determined jointly by the weighted phase space and the anharmonic scattering matrix elements. Consequently, a larger weighted phase space alone does not necessarily imply a proportionally larger phonon scattering rate.

To account for higher-order anharmonicity, four-phonon scattering processes were explicitly incorporated into the lattice thermal conductivity calculations using the \textsc{FourPhonon} package \cite{Han2022}. The intrinsic four-phonon scattering rates were evaluated within fourth-order many-body perturbation theory using Fermi's golden rule following the formalism developed by Feng \textit{et al.} \cite{Feng2016,Feng2017}. The implementation explicitly considers all symmetry-allowed $1\leftrightarrow3$ and $2\leftrightarrow2$ phonon scattering processes satisfying both crystal momentum and energy conservation. The temperature dependence of the scattering rates naturally arises from the Bose-Einstein occupation factors associated with each scattering channel.

The total intrinsic phonon scattering rate was obtained through Matthiessen's rule,
\begin{equation}
	\frac{1}{\tau_{\mathbf{q}j}}
	=
	\frac{1}{\tau_{\mathbf{q}j}^{(3)}}
	+
	\frac{1}{\tau_{\mathbf{q}j}^{(4)}},
\end{equation}
where $\tau_{\mathbf{q}j}^{(3)}$ and $\tau_{\mathbf{q}j}^{(4)}$ denote the relaxation times arising from three-phonon and four-phonon scattering, respectively. The resulting total scattering rates were subsequently employed within the iterative solution of the linearized Boltzmann transport equation to evaluate the intrinsic lattice thermal conductivity.

Finally, the intrinsic lattice thermal conductivity was calculated over the temperature range of 250-1000 K by iteratively solving the $l$BTE using both the conventional three-phonon formalism and the combined three-phonon plus four-phonon scattering framework. To elucidate the microscopic mechanisms governing thermal transport in monolayer beryllene, mode-resolved phonon lifetimes, scattering rates, weighted scattering phase space, phonon mean free paths, and accumulated lattice thermal conductivity were systematically analyzed. Furthermore, the branch-resolved contributions of the flexural acoustic (ZA), transverse acoustic (TA), and longitudinal acoustic (LA) phonons were evaluated to quantify the influence of higher-order anharmonicity on the dominant heat-carrying phonons.

\section{\label{sec:result}Results and Discussion\protect}
\subsection{\label{sec:phonondos}Structural Configuration and Phonons}
\begin{figure}[h]
	\includegraphics[height=6cm, keepaspectratio]{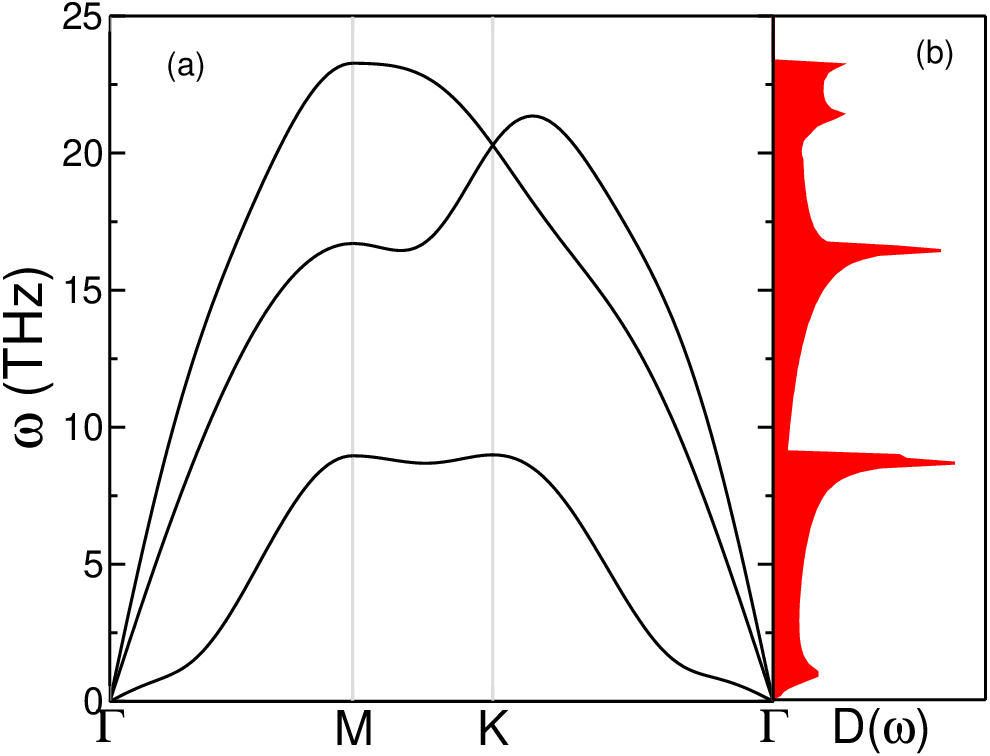}% Here is how to import EPS art
	\caption{(a) Phonon dispersion and (b) phonon density of states (DOS) for monolayer beryllene, calculated using density functional perturbation theory (DFPT). The absence of optical modes, with only acoustic phonon branches present, reflects the monatomic nature of the beryllene monolayer.}
	\label{fig:band_dos}
\end{figure}

The monolayer structure of beryllene exhibits a distinctive hexagonal planar configuration with six first nearest neighbors, characterized by a lattice constant of 2.13 Å. This structure belongs to the P6/mmm space group and D$_{6h}$ point group symmetry, consistent with earlier theoretical predictions and experimental observations \cite{Chahal2023, beryllene_sun20}. The phonon dispersion relations, plotted along the high-symmetry directions $\Gamma\rightarrow$M$\rightarrow$K$\rightarrow\Gamma$ in Fig. \ref{fig:band_dos}(a), reveal a maximum phonon frequency of 23.4 THz at the $\Gamma$ point, which is consistent with prior studies \cite{beryllene_sun20, Li2023} confirming the validity and accuracy of our computational methods.

In Fig. \ref{fig:band_dos}(b), we present the phonon density of states (PDOS) for monolayer beryllene, plotted against the angular frequency. Since the precision of phonon dispersion is critical for accurate thermal conductivity calculations, we have systematically optimized the phonon frequencies by adjusting the kinetic energy cutoffs for the pseudopotential employed in this study. Due to the monatomic nature of beryllene, the phonon spectrum consists exclusively of acoustic modes, specifically the out-of-plane ZA mode and the in-plane TA and LA modes. Importantly, the absence of any soft phonon modes indicates the dynamical stability of $\alpha$-beryllene. The out-of-plane ZA mode displays a characteristic quadratic behavior, while the in-plane TA and LA modes exhibit cubic behavior, analogous to the behavior observed in graphene \cite{Gu2015}. The quadratic dispersion of the ZA mode can be attributed to the layer bending stiffness, a phenomenon well-explained by the elastic plate theory, while the cubic behavior of the in-plane modes is a result of in-plane stretching vibrations \cite{Zabel2001}. Additionally, the sharp features in the PDOS correspond to saddle points at the M and K points in the Brillouin zone, showing the complex vibrational landscape of this material.

\subsection{\label{subsec:cond}Thermal Conductivity and Temperature Dependence}
\begin{figure}[h]
	\includegraphics[height=6cm,keepaspectratio]{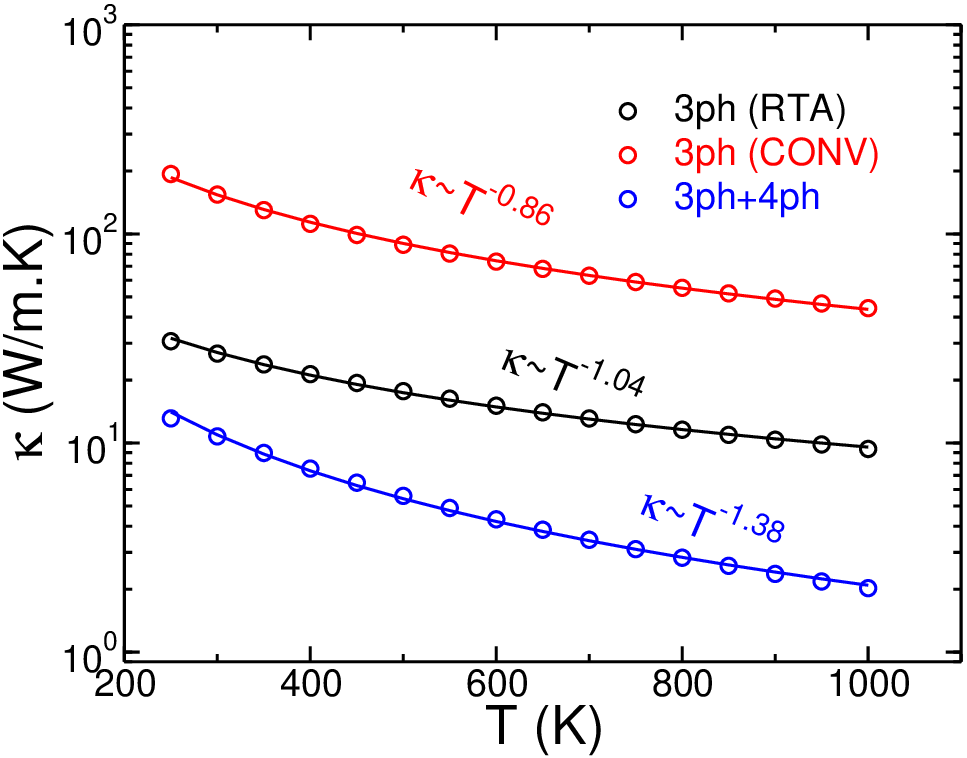}% Here is how to import EPS art
	\caption{Temperature-dependent thermal conductivity of monolayer beryllene, calculated using the linearized Boltzmann transport equation. The three-phonon thermal conductivities are calculated using the relaxation time approximation (black circles) and an iterative solver (red circles). The thermal conductivity calculated by including four-phonon interactions are shown using blue circles.}
	\label{fig:kappa_vs_T}
\end{figure}
We calculate the lattice thermal conductivity of $\alpha$-beryllene by solving the phonon Boltzmann transport equation (BTE) using first-principles harmonic, third-order, and fourth-order interatomic force constants. Fig. \ref{fig:kappa_vs_T} compares the thermal conductivity obtained within the relaxation time approximation (3ph-RTA), the fully converged solution of the linearized BTE considering only three-phonon scattering (3ph-CONV), and the solution including both three- and four-phonon scattering processes (3ph+4ph). At room temperature, the thermal conductivity predicted by the converged three-phonon solution is approximately 154.3 W/(m$\cdot$K), substantially larger than the corresponding RTA value ($\sim$25 W/(m$\cdot$K)), which indicates that normal phonon scattering plays a dominant role in heat transport. Because normal processes conserve crystal momentum, they redistribute phonons without directly relaxing the heat current, leading to a significant enhancement of thermal conductivity in the exact solution of the BTE compared with the relaxation time approximation. Similar behaviour has been reported for high-thermal-conductivity crystals such as diamond, graphene and cubic boron arsenide, where momentum-conserving normal processes produce hydrodynamic-like phonon transport and cause the RTA to substantially underestimate the lattice thermal conductivity \cite{Cepellotti2015, Fugallo2014, Lindsay2013}.

When intrinsic four-phonon scattering is included, the thermal conductivity decreases dramatically over the entire temperature range. At 300 K, the calculated thermal conductivity is reduced to 10.7 W/(m$\cdot$K), corresponding to nearly an order-of-magnitude reduction compared with the converged three-phonon result. This pronounced suppression indicates that four-phonon scattering provides significant resistive channels for heat transport in $\alpha$-beryllene and therefore must be included in the estimation of thermal conductivity. Strong four-phonon effects have been reported for several materials with strong anharmonicity, including graphene, diamond, and boron arsenide, where higher-order phonon interactions substantially reduce the thermal conductivity obtained from three-phonon calculations alone \cite{Feng2017, Ravichandran2018, Jain2024, Sharma2026, Chaudhuri2024}. However, the reduction in $\alpha$-beryllene is significantly higher than that reported in other 2D materials. 

To place the present results in a broader context, Table I compares the influence of four-phonon scattering on the lattice thermal conductivity of representative materials reported using first-principles Boltzmann transport calculations. Although the absolute thermal conductivity varies substantially owing to differences in bonding, dimensionality, and phonon spectra, the relative reduction upon including four-phonon scattering provides a useful measure of the importance of higher-order anharmonicity. Four-phonon interactions produce only modest reductions in conventional semiconductors such as diamond and silicon at room temperature, whereas much stronger effects have been reported for materials with highly restricted three-phonon scattering channels or pronounced anharmonicity, including graphene, boron arsenide, MoS$_2$, and single-walled carbon nanotubes \cite{Feng2017, Ravichandran2018, Jain2024, Sharma2026, Chaudhuri2024}.

As shown in Table I, $\alpha$-beryllene exhibits one of the strongest reported suppressions of lattice thermal conductivity due to explicit four-phonon scattering, with the room-temperature conductivity decreasing from 154.3 to 10.7 W/(m$\cdot$K), corresponding to a reduction of approximately $\sim$93\%. This reduction exceeds those reported for most benchmark crystalline materials considered using comparable first-principles approaches. Such a pronounced suppression signifies that higher-order anharmonicity plays an exceptionally important role in heat transport in $\alpha$-beryllene and confirms that three-phonon scattering alone is insufficient for quantitatively reliable predictions of its intrinsic lattice thermal conductivity. The strong sensitivity to four-phonon scattering motivates the detailed mode-resolved analysis presented in the following sections.

The temperature dependence of the thermal conductivity follows a power-law behaviour throughout the investigated range of 250-1000 K. The converged three-phonon solution exhibits a relatively weak dependence, following approximately $\kappa \sim T^{-0.86}$, whereas the RTA solution scales as $\kappa \sim T^{-1.04}$, close to the classical inverse-temperature behaviour expected when Umklapp scattering dominates thermal resistance \cite{Callaway1959, Broido2007}. In contrast, including four-phonon scattering leads to a noticeably stronger temperature dependence, $\kappa \propto T^{-1.38}$. The steeper decay originates from the rapid increase in the available four-phonon scattering phase space with temperature, which causes four-phonon scattering rates to increase more rapidly than conventional three-phonon processes. As a consequence, four-phonon scattering becomes increasingly important at elevated temperatures and accelerates the reduction of thermal conductivity beyond that predicted by three-phonon interactions alone \cite{Feng2017, Chaudhuri2024}.

It should be emphasized that $\alpha$-beryllene is a metallic two-dimensional material. Consequently, the total thermal conductivity consists of both lattice (phononic) and electronic contributions, $\kappa_{\mathrm{tot}}=\kappa_{\mathrm{L}}+\kappa_{\mathrm{e}}$. Our work focuses exclusively on the intrinsic lattice thermal conductivity. Therefore, the reported values quantify only the phonon-mediated component of heat transport. Further, the present calculations are performed within the conventional perturbative framework using interatomic force constants obtained at 0 K. Consequently, temperature-induced phonon renormalization arising from higher-order anharmonicity, which can substantially modify phonon frequencies, lifetimes, and thermal conductivity in strongly anharmonic materials, is not explicitly considered \cite{Han2022PRL}. Such a comprehensive first-principles treatment remains an important direction for future investigation.

\begin{table*}[t]
	\caption{Comparison of representative first-principles studies reporting the influence of four-phonon scattering on lattice thermal conductivity in low-dimensional and bulk crystalline materials. The lattice thermal conductivities are obtained by including only three-phonon scattering ($\kappa_{3}$) and combined three- and four-phonon scattering ($\kappa_{3+4}$). The percentage reduction is calculated as
		$\left(1-\kappa_{3+4}/\kappa_{3}\right)\times100$.}
	\label{tab:comparison}
	\begin{ruledtabular}
		\begin{tabular}{lcccccl}
			Material & $T$ (K) &
			$\kappa_{3}$ &
			$\kappa_{3+4}$ &
			Reduction &
			Reference \\
			&
			&
			(W\,m$^{-1}$K$^{-1}$) &
			(W\,m$^{-1}$K$^{-1}$) &
			(\%) &
			\\
			\hline
			Diamond          & 300  & $\sim2280$ & $\sim2000$ & 12.3 &  \cite{Feng2017} \\
			Diamond          & 1000 & 658.0       & 519.0      & 21.1 &  \cite{Feng2017}\\
			
			Graphene         & 300  & 3340        & 1839       & 44.9 &  \cite{Xu2025}\\
			
			TTH Carbon (x)   & 300  & 699.5       & 372.6      & 46.7 &  \cite{Jia2025}\\
			TTH Carbon (y)   & 300  & 952.6       & 649.3      & 31.8 &  \cite{Jia2025}\\
			
			Silicon          & 300  & 150         & 130        & 13.3 &  \cite{Feng2017} \\
			
			BAs              & 300  & $\sim2279$  & 1436       & 37.0 & \cite{Feng2017}\\
			BAs              & 1000 & 824.8       & 193.9      & 76.5 &  \cite{Feng2017}\\
			
			TiB$_2$          & 300  & $\sim352$   & $\sim228$  & 35.2 & \cite{Sharma2026} \\
			
			SWCNT            & 300  & $\sim9000$  & $\sim4000$ & 55.6 & \cite{Jain2024} \\
			
			MoS$_2$          & 300  & $\sim116$   & $\sim25$   & 78.4 &  \cite{Chaudhuri2024} \\
			
			\textbf{$\alpha$-Beryllene (This work)}
			& \textbf{300}
			& \textbf{154.3}
			& \textbf{10.7}
			& \textbf{$\sim$93}
			& \textbf{This work} \\
		\end{tabular}
	\end{ruledtabular}
\end{table*}

\subsection{\label{subsec:mechanisms}Mechanisms Underlying Phonon Thermal Transport}
In this section, we provide an in-depth analysis of the phonon-mediated heat transport mechanisms in $\alpha$-beryllene. We begin by examining the normalized accumulated thermal conductivity as a function of mean free path (MFP) at room temperature, along with the accumulated thermal conductivity as a function of phonon angular frequency. These analyses shed light on the contributions of different phonon modes to the overall thermal conductivity. To further elucidate the underlying mechanisms, we conduct a thorough investigation into the phonon linewidths and weighted phase space, with particular emphasis on the anisotropic distribution of thermal conductivity across various acoustic modes. %This detailed examination allows us to uncover the intricate interplay between phonon scattering processes and thermal transport, offering new insights into the fundamental factors that govern heat conduction in $\alpha$-beryllene.
\begin{figure}[h]
	\includegraphics[height=6cm,keepaspectratio]{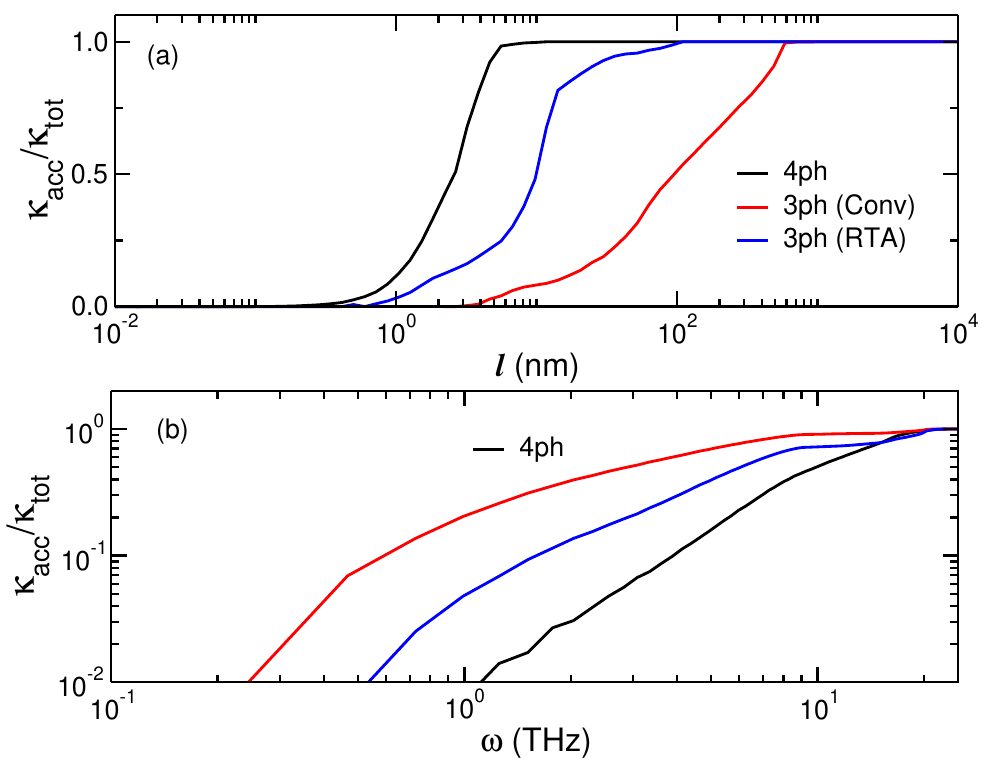}% Here is how to import EPS art
	\caption{(a) Normalized accumulated thermal conductivity of monolayer beryllene at room temperature as a function of phonon mean free path. (b) Accumulated thermal conductivity as a function of phonon angular frequency. Accumulation of thermal conductivity for different solutions of the BTE is shown. It can be seen that the cumulative thermal conductivity saturates $\sim$11 nm for the solution with fourthorder interaction included.}
	\label{fig:acc_kappa}
\end{figure}
\subsubsection{\label{subsubsec:mechanisms-part1}Spectral Decomposition of Accumulated Thermal Conductivity}
\begin{figure}[h]
	\includegraphics[height=6cm,keepaspectratio]{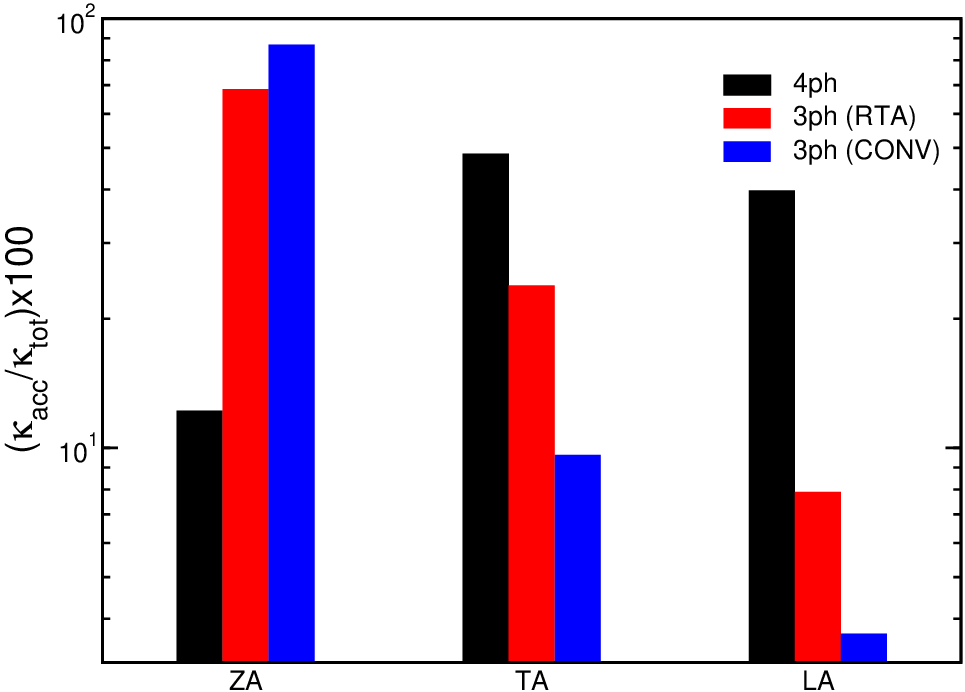}% Here is how to import EPS art
	\caption{Percentage contribution to the total thermal conductivity as a function of different phonon branches. It can be observed that the out-of-plane ZA phonon contributes significantly (~80\%) to the thermal conductivity in the case of solution with only three-phonon scattering considered. Upon including the four-phonon scattering, the contribution of the ZA mode becomes the least, and the contributions due to different branches become more even.}
	\label{fig:contribution}
\end{figure}
Fig. \ref{fig:acc_kappa}(a) shows the normalized accumulated lattice thermal conductivity as a function of phonon mean free path (MFP) at room temperature for the three scattering models considered: the full four-phonon calculation (3ph+4ph), the iterative solution of the three-phonon Boltzmann transport equation (3ph Conv.), and the three-phonon relaxation time approximation (3ph RTA). The inclusion of four-phonon scattering shifts the accumulation curve markedly toward shorter mean free paths. This shows that fourphonon interactions substantially reduce the transport lengths of heat-carrying phonons. In the fourphonon based calculation, nearly the entire thermal conductivity is accumulated within only a few nanometers, whereas both three-phonon calculations predict a much broader MFP distribution extending to considerably longer length scales. Likewise, the median heat-carrying phonons (corresponding to 50\% accumulated thermal conductivity) possess significantly shorter MFPs when four-phonon scattering is included. These observations indicate that higher-order anharmonicity efficiently suppresses long-lived phonons, thereby reducing the characteristic transport length governing intrinsic thermal conduction \cite{Feng2016,Lindsay2010}.

The pronounced shift toward shorter MFPs has important implications for thermal transport in finite-sized samples. Since the intrinsic phonon MFPs are substantially reduced by four-phonon scattering, the sample dimensions required for boundary scattering to appreciably suppress thermal conductivity become correspondingly smaller. Consequently, calculations neglecting four-phonon processes are expected to overestimate both the intrinsic thermal conductivity and the characteristic transport length scale relevant to nanoscale heat transport \cite{Feng2016,Fugallo2014}.

To further elucidate the spectral origin of heat conduction, Fig. \ref{fig:acc_kappa}(b) presents the normalized accumulated thermal conductivity as a function of phonon frequency. The three-phonon calculations indicate that most of the thermal conductivity originates from low-frequency acoustic phonons. However, once four-phonon scattering is incorporated, the accumulation curve shifts toward higher frequencies, implying that the lowest-frequency phonons undergo considerably stronger higher-order anharmonic scattering. As a result, their contribution to thermal transport is substantially reduced, while intermediate-frequency acoustic phonons account for a larger fraction of the total thermal conductivity. The 3ph-RTA results lie between the iterative three-phonon solution and the four-phonon calculation. This is due to the fact that the relaxation time approximation neglects the redistribution of nonequilibrium phonon populations arising from momentum-conserving normal processes captured by the iterative solution \cite{Li2014ShengBTE}. These results signify that four-phonon scattering modifies not only the magnitude of thermal conductivity but also the spectral distribution of the dominant heat-carrying phonons.

The branch-resolved thermal conductivity shown in Fig. \ref{fig:contribution} provides further insight into the microscopic origin of this behavior. Within both three-phonon calculations, the ZA branch dominates thermal transport, contributing substantially more than the in-plane TA and LA branches. Such behavior is characteristic of suspended two-dimensional crystals, where the quadratic dispersion and large phonon population of flexural modes lead to exceptionally high contributions to lattice thermal conductivity when only three-phonon scattering is considered \cite{Lindsay2010,Lindsay2014,Xie2016}. In contrast, the inclusion of four-phonon scattering dramatically suppresses the ZA contribution, while the reductions in the TA and LA branches are comparatively less severe. Consequently, thermal transport becomes considerably more evenly distributed among the three acoustic branches.

The strong suppression of ZA-mediated heat transport originates from the additional scattering channels introduced by four-phonon interactions, which substantially enhance the total scattering rates of low-frequency ZA phonons. Although flexural phonons are subject to symmetry-based selection rules that restrict several three-phonon scattering channels in ideal two-dimensional crystals, four-phonon processes introduce additional symmetry-allowed scattering pathways. These higher-order interactions substantially increase the scattering rates of low-frequency ZA phonons, thereby shortening their lifetimes and mean free paths and reducing their contribution to the lattice thermal conductivity \cite{Lindsay2010,Feng2016,Feng2017}. Similar behavior has been reported for graphene, where explicit inclusion of four-phonon scattering significantly lowers the thermal conductivity through enhanced scattering of flexural acoustic phonons \cite{Feng2016,Feng2017}.

\subsubsection{\label{subsubsec:mechanisms-part2}Phonon Scattering and Weighted Phase Space Analysis}
\begin{figure}[h]
	\includegraphics[height=6cm,keepaspectratio]{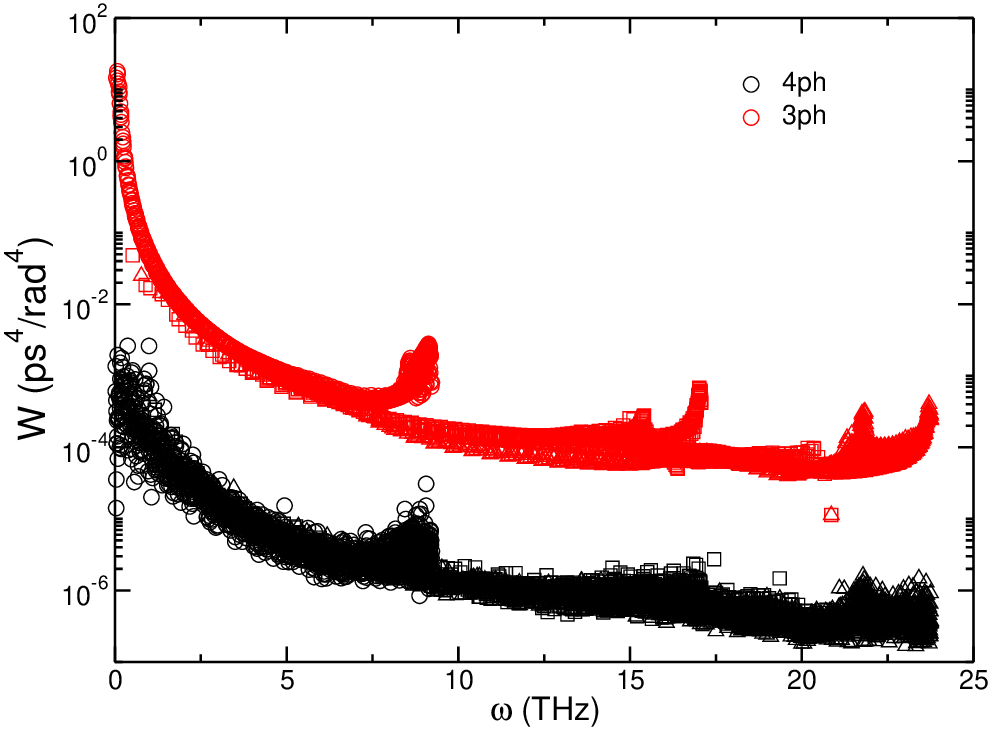}% Here is how to import EPS art
	\caption{Weighted scattering phase space for three-phonon and four-phonon processes as a function of angular frequency. The four-phonon phase space is generally smaller in magnitude than the three-phonon phase space, indicating that the pronounced enhancement of four-phonon scattering rates cannot be attributed to phase space alone and also reflects the underlying anharmonic interaction strengths.}
	\label{fig:phase_sapce}
\end{figure}

To elucidate the microscopic origin of the substantial reduction in lattice thermal conductivity upon inclusion of four-phonon interactions, we analyze the weighted scattering phase space together with the mode-resolved phonon scattering rates. The weighted phase space, $W(\omega)$, quantifies the number of energy- and momentum-conserving scattering channels available to a phonon mode while accounting for the phonon occupation factors entering the scattering probability \cite{Li2014ShengBTE,Togo2015}. Although the weighted phase space provides a useful measure of the available scattering channels, the actual phonon scattering rates are determined jointly by the available phase space and the corresponding anharmonic interaction matrix elements through Fermi's golden rule \cite{Feng2016,Feng2017}.

Fig. ~\ref{fig:phase_sapce} compares the weighted phase space associated with three-phonon and four-phonon scattering processes. In both cases, the weighted phase space is largest in the low-frequency acoustic region and generally decreases with increasing frequency, although several frequency-dependent features are visible in the intermediate- and high-frequency regimes. Since $\alpha$-beryllene has a single-atom primitive cell, all phonon modes belong to the three acoustic branches, namely ZA, TA, and LA. Although the frequency dependence of the three-phonon and four-phonon weighted phase spaces is qualitatively similar, their absolute magnitudes should not be compared directly because the corresponding quantities are defined within different scattering formalisms and have different dimensionality. The enhanced four-phonon scattering rates observed particularly for the low-frequency ZA modes therefore cannot be attributed to the weighted phase space alone, but result from the combined effects of the available scattering channels and the corresponding anharmonic interaction matrix elements.

\begin{figure}[h]
	\includegraphics[height=6cm,keepaspectratio]{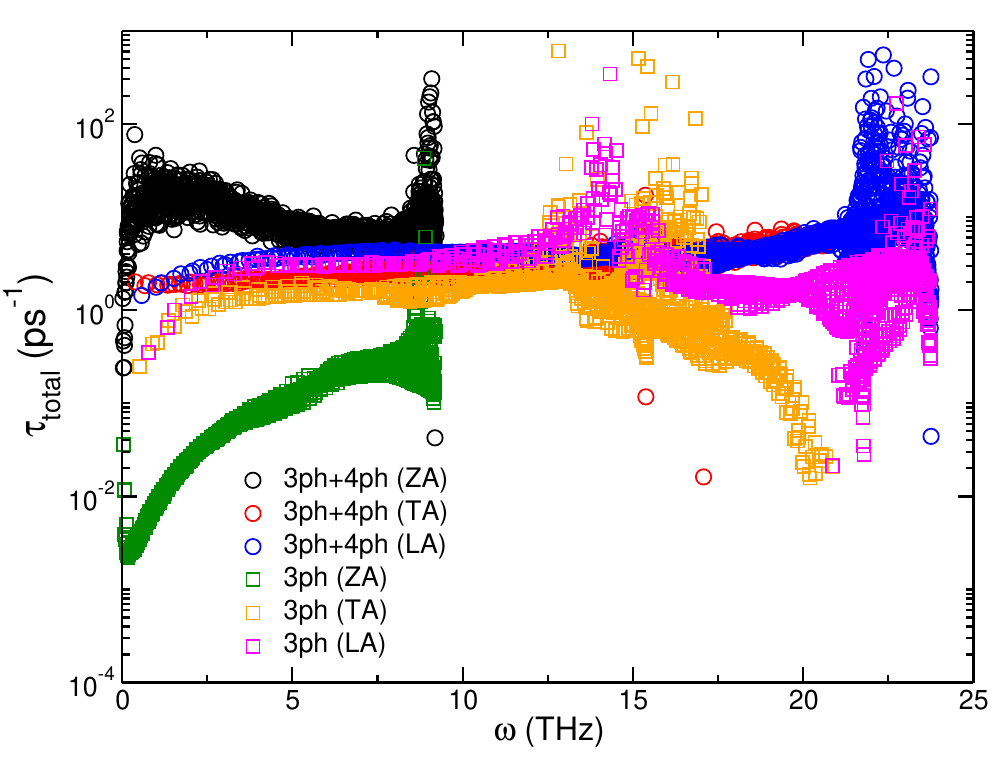}% Here is how to import EPS art
	\caption{Phonon scattering rates (in ps$^{-1}$) of the different phonon branches in $\alpha$-beryllene. The results compare three-phonon scattering alone (squares) with the combined three- and four-phonon scattering rates (circles). The pronounced enhancement of the ZA scattering rate at low frequencies upon inclusion of four-phonon interactions signifies the strong influence of higher-order anharmonicity on flexural phonons.}
	\label{fig:anharmonic_scattering}
\end{figure}

The influence of the additional four-phonon scattering processes is more clearly reflected in the phonon scattering rates shown in Fig. \ref{fig:anharmonic_scattering}. In the three-phonon calculations, the ZA branch exhibits comparatively weak scattering at low frequencies, resulting in long phonon lifetimes and making ZA phonons the dominant heat carriers. Upon inclusion of four-phonon interactions, however, the scattering rates of low-frequency ZA phonons increase markedly, with the largest enhancement occurring below approximately 5~THz. In comparison, the TA and LA branches exhibit substantially smaller changes, although noticeable increases in their scattering rates are observed throughout the acoustic frequency range. The pronounced enhancement of the ZA scattering rate therefore indicates that higher-order anharmonicity predominantly affects the flexural phonons that carry a large fraction of the heat within the three-phonon approximation.

The strong enhancement of ZA scattering can be understood in terms of the symmetry of the planar monolayer. For an ideal two-dimensional crystal possessing reflection symmetry with respect to the plane of the monolayer, selection rules restrict three-phonon processes involving an odd number of flexural ZA phonons, thereby reducing their scattering rates and resulting in comparatively long ZA phonon lifetimes \cite{Lindsay2010,bonini2012acoustic}. Four-phonon processes remain subject to the same symmetry constraints but provide additional symmetry-allowed scattering channels that are absent from a purely three-phonon description \cite{Feng2016,Feng2017}. Consequently, the total scattering rate of low-frequency ZA phonons can increase substantially when four-phonon interactions are included, leading to shorter phonon lifetimes and a marked suppression of their contribution to the lattice thermal conductivity.

The microscopic scattering behavior is consistent with the thermal transport characteristics presented in Fig. \ref{fig:kappa_vs_T} and Fig. \ref{fig:acc_kappa}. The enhanced four-phonon scattering shortens phonon lifetimes and mean free paths, shifts the accumulated thermal conductivity toward significantly shorter transport lengths, and substantially suppresses the dominant ZA contribution. As a result, the room-temperature lattice thermal conductivity decreases from approximately 154.3 W/(m$\cdot$K) obtained from the iterative three-phonon solution to about 10.7 W/(m$\cdot$K) when four-phonon scattering is explicitly included, corresponding to a reduction of approximately 93\%. Furthermore, the temperature dependence changes from $\kappa \sim T^{-0.86}$ in the three-phonon calculation to $\kappa \sim T^{-1.38}$ after incorporating four-phonon scattering. The stronger temperature dependence reflects the increasing importance of four-phonon scattering with temperature, arising from the temperature dependence of the phonon occupation factors and the corresponding increase in four-phonon scattering rates. Similar behavior has been reported in graphene, diamond, boron arsenide, and other high-thermal-conductivity materials, where four-phonon scattering provides an important intrinsic resistance to heat transport, particularly at room temperature and above \cite{Feng2016,Feng2017,Lindsay2019}.

Overall, the weighted phase-space analysis and the branch-resolved scattering rates establish a consistent microscopic picture of thermal transport in monolayer beryllene. The inclusion of four-phonon interactions introduces additional symmetry-allowed scattering pathways and substantially enhances the scattering of low-frequency ZA phonons, resulting in shorter phonon lifetimes and reduced mean free paths. The resulting suppression of ZA-mediated heat transport shifts a larger fraction of the thermal conductivity toward the in-plane TA and LA branches. These effects collectively explain the pronounced reduction in lattice thermal conductivity obtained from the full four-phonon calculations and show that explicit inclusion of higher-order anharmonicity is essential for quantitatively reliable predictions of thermal transport in strongly anharmonic two-dimensional materials.

\section{\label{sec:conclusions}Conclusion\protect}
In conclusion, we calculate the thermal conductivity of $\alpha$-beryllene using both three phonon and four phonon approximations. We demonstrate that explicit inclusion of four-phonon interactions reduces the predicted room-temperature lattice thermal conductivity from 154.3 W/(m·K) to 10.7 W/(m·K). This suppression of approximately 93\% represents one of the most extreme renormalizations observed among two-dimensional materials to date. The microscopic origin of this transport breakdown stems from a massive enhancement in the scattering rates of low-frequency flexural (ZA) phonons. We find that higher-order anharmonicity selectively curtails the lifetimes and mean free paths of these originally long-lived modes. Consequently, the primary channels for heat conduction are fundamentally redistributed away from flexural modes and toward shorter-lived in-plane acoustic phonons. This shift is strictly corroborated by a steepening of the thermal conductivity temperature dependence, which transitions from $\kappa \propto T^{-0.86}$ to $\kappa \propto T^{-1.38}$ when four-phonon scattering is incorporated. Ultimately, these findings position $\alpha$-beryllene as a benchmark system for four-phonon-dominated thermal transport.

\section*{SUPPLEMENTARY MATERIAL}
The supplementary material contains convergence tests for (S1) Plane-wave energy cutoff, (S2) Four-phonon thermal conductivity, (S3) Sampling of four phonon scattering processes, and (S4) \textit{q}-point mesh.
\begin{acknowledgments}
	The authors acknowledge the Computer Center of IIT Jodhpur for providing computing resources that have contributed to the research results reported in this paper.
\end{acknowledgments}
%\section*{CRediT author statement}

%\textbf{Sapta Sindhu Paul Chowdhury}: Conceptualization, Methodology, Software, Investigation, Formal analysis,  Writing - Original draft preparation, Visualization; \textbf{Santosh Mogurampelly}: Conceptualization, Methodology, Writing - Review \& Editing, Supervision.
\section*{Conflicts of interest}
The authors have no conflicts to disclose.
\section*{Data Availability Statement}
The data that support the findings of this study are available from the corresponding author upon reasonable request. 
%\section*{Data Availability}
%The datasets generated and/or analyzed during the current study are not publicly available due to their large size, but are available from the corresponding authors on reasonable request.
%\section*{Funding Statement}
%There is no funding for this study.
%\nocite{*}
\bibliography{merged}% Produces the bibliography via BibTeX.
\end{document}